\documentclass[12pt]{article}
\usepackage{amsfonts}
\usepackage{amssymb}
\usepackage{amsbsy}
\usepackage{latexsym}
\usepackage{wrapfig}
\usepackage[dvips]{graphicx}
\evensidemargin=\oddsidemargin
\usepackage{epsfig}
\newcommand{\beq}{\begin{equation}}
\newcommand{\eeq}{\end{equation}}
\newcommand{\bea}{\begin{eqnarray}}
\newcommand{\eea}{\end{eqnarray}} 
\newcommand{\ba}{\begin{array}}
\newcommand{\ea}{\end{array}} 
\newcommand{\ds}{\displaystyle} 
\newcommand{\bx}{{\bf x}}
\newcommand{\bp}{{\bf p}}
\newcommand{\by}{{\bf y}}
\newcommand{\bk}{{\bf k}}
\newcommand{\bn}{{\bf n}}
\newcommand{\brho}{{\boldsymbol\rho}}

\newcommand{\tphi}{\tilde\phi}

\newcommand{\p}{\partial}
\begin{document}
\begin{flushright}
hep-th/0204110\\ 
HIP-2002-12/TH \\ 
\end{flushright}
\vspace*{3mm}
\begin{center}
{\Large {\bf Trans-Planckian Effects in Inflationary Cosmology} \\[.1cm] 
{\bf and the Modified Uncertainty Principle}\\} 
\vspace*{12mm} 
{\bf S. F. Hassan
\footnote{{\tt e-mail: fawad.hassan@helsinki.fi} (After August 2002, 
{\tt fawad@physto.se})}
and Martin S. Sloth
\footnote{\tt e-mail: martin.sloth@helsinki.fi}\\}
\vspace{3mm}
{\it Helsinki Institute of Physics\\[.1cm] 
P.O. Box 64, FIN-00014 University of Helsinki, Finland} 
\vspace{1cm}
\begin{abstract}
\noindent There are good indications that fundamental physics gives
rise to a modified space-momentum uncertainty relation that implies
the existence of a minimum length scale. We implement this idea in the
scalar field theory that describes density perturbations in flat
Robertson-Walker space-time. This leads to a non-linear time-dependent
dispersion relation that encodes the effects of Planck scale physics
in the inflationary epoch. Unruh type dispersion relations naturally
emerge in this approach, while unbounded ones are excluded by the
minimum length principle. We also find red-shift induced modifications
of the field theory, due to the reduction of degrees of freedom at
high energies, that tend to dampen the fluctuations at trans-Planckian
momenta. In the specific example considered, this feature helps
determine the initial state of the fluctuations, leading to a flat
power spectrum.
\end{abstract}
\end{center}
\setcounter{footnote}{0}
\baselineskip16pt

\section{Introduction}

According to the inflationary scenario
\cite{Guth:1980zm},\cite{Linde:1981mu},\cite{Albrecht:1982wi} in the
past the universe has undergone a period of rapid expansion during
which minute quantum fluctuations have been magnified to cosmic sizes.
This gives rise to the inhomogeneities that are observed today, in the
form of anisotropies in the cosmic microwave background radiation
(CMBR) and large scale structures, superimposed on a homogeneous and
isotropic background. To describe the primordial quantum fluctuations,
it has so far sufficed to use a prototype free scalar field theory
with a linear dispersion relation in the expanding background
\cite{Mukhanov:1990me}. The prototype scalar field describes the
tensor and scalar metric (or inflaton) fluctuations. The success of
the inflationary scenario rests on its ability to explain not only the
homogeneity of the background, but also the characteristics of the
inhomogeneities superimposed upon it.

However, as pointed out in \cite{Brandenberger:1999sw}, in many
versions of inflation, most notably chaotic inflation, the period of
inflation lasts for so long that the co-moving scales of cosmological
interest today correspond to quantum fluctuations at the beginning of
inflation with physical wavenumbers larger than the Planck mass, or
equivalently, with physical wavelengths smaller than the Planck
length. It is clear that for such fluctuations the effects of Planck
scale physics cannot be ignored and the use of a field theory with a
linear dispersion relation is not fully justified. This has been
referred to as the trans-Planckian problem of inflation. It opens up
the interesting possibility of searching for the imprints of Planck
scale physics in the large scale structure of the universe.

Of late, it has been customary to assume that the effects of Planck
scale physics can be incorporated in the theory of cosmological
perturbations by using a time dependent non-linear dispersion relation
in the free field theory. In the absence of a first principle
approach, in practice these dispersion relations are chosen in an {\it
ad hoc} manner. In the context of cosmology, the study of the
consequences of such modifications was taken up in
\cite{Martin:2000xs},\cite{Brandenberger:2000wr}. The modifications
introduced were inspired by earlier work in the context of Hawking
radiation \cite{Unruh:1995},\cite{Corley:1996ar}. Subsequently, 
various aspects of this issue were further investigated in 
\cite{Martin:2000bv} - \cite{Danielsson:2002kx}.

A direct derivation of the trans-Planckian effects from a fundamental
theory, {\it e.g.}, string theory is not yet possible. However, Planck
scale physics does seem to have a robust characteristic feature, which
is the existence of a minimum length scale. For a review, see
\cite{Garay:1994en}. In particular, in perturbative string theory,
this feature has been known for some time
\cite{Gross:1987ar},\cite{Amati:1988tn}. There, the existence of a
minimum length uncertainty is essentially due to the fact that as one
increases the energy of a string probe beyond the Planck mass, the
string can get excited and develop a non-zero extension \cite{witten}.
The existence of a minimum length uncertainty can be related to a
modification of the standard space-momentum canonical commutation
relations. For an implementation of this idea, see
\cite{Kempf:1994su},\cite{Kempf:1996nk}. Thus, one way of
incorporating Planck scale physics in a field theory is to reformulate
the theory such that it is consistent with the minimum length
uncertainty relation. This is the strategy that we will follow in this
paper. Earlier an attempt in this direction was made in
\cite{Kempf:2000ac}, followed by further analysis in
\cite{Kempf:2001fa},\cite{Easther:2001fz}. This resulted in a rather
complicated theory which cannot easily be compared with other models
and in which the identification of physical variables is not
straightforward. In this paper, we follow a different approach to
implement the minimum length uncertainty principle in a field theory
in the expanding universe. This results in a theory with a transparent
physical content. It has a non-linear time-dependent dispersion
relation which is no longer arbitrary, but is determined by the
modified commutation relations. We show that this naturally leads to
Unruh type dispersion relations. Besides this, the theory contains a
trans-Planckian damping term due to the reduction of the degrees of
freedom at high momenta. This tends to suppress the fluctuations at
momenta sufficiently above the Planck scale, although the actual
amount of damping depends on the form of the dispersion relation. It
is interesting to note that there are independent arguments for
trans-Planckian damping due to quantum gravity effects
\cite{DeWitt:64},\cite{Padmanabhan:jp}.

The paper is organized as follows: In section 2 we describe a class of
modified $\bx,\bp$ commutation relations that lead to a minimum length
uncertainty principle in flat space-time. Then we construct a free
scalar field theory based on this and interpret the modification in
terms of higher derivative corrections. We also discuss the modified
relation between wavelength and wavenumber. In section 3, we
generalize the notion of modified commutators and the associated
minimum length uncertainty to the flat Robertson-Walker space-time. In
section 4, we apply this to a scalar field theory in flat
Robertson-Walker space-time and obtain its modified equation of
motion. We discuss the resulting non-linear and time-dependent
dispersion relations which are now constrained by the minimum length
principle. The time evolution of wavelengths is also discussed. In
section 5, we analyze a new feature of the modified field theory which
tends to dampen the fluctuations at trans-Planckian energies. For the
dispersion relation used there, this feature allows us to determine
the initial state, and leads to a scale invariant power spectrum for
the fluctuations. The issue of back reaction is also discussed and the
exact solution in the trans-Planckian region is presented. In
subsection 5.3, we discuss the issue of the initial state dependence
of the power spectrum and argue that the main criterion is whether the
effective frequency at the initial time is real or imaginary. This is
related to, but not the same as, the violation of adiabaticity during
time evolution. Throughout this paper we assume de Sitter inflation so
that the Hubble parameter is constant.

\section{The minimum length uncertainty principle}

Conventional quantum mechanics allows us to use trans-Planckian
momenta to probe distances shorter than a Planck length, ignoring the
issue of gravitational stability of matter. The problem can be avoided
if one assumes that nature admits a description, at least in an
effective sense, in terms of a theory with a minimum length scale.
Such a minimum length scale follows from (or results in) a modified
uncertainty relation. There are indications that fundamental physics
\cite{Garay:1994en}, {\it e.g.}, string theory effects
\cite{Gross:1987ar},\cite{Amati:1988tn},\cite{witten} {\it effectively}
modify the usual space-momentum uncertainty relation to
\beq
\Delta x \geq \frac{1}{2}\,(\,\frac{1}{\Delta p}\,+\, l_s^2\, \Delta p 
\,+\,\cdots\, )\,.
\label{guc}
\eeq 
This implies the existence of a lower bound on space uncertainty
$(\Delta x)_{min}\sim l_s$. The scale $l_s$ is a minimum length that
we take as the Planck length, although it could also be some other
related scale. Our goal in this paper is to explore the implications
of such a minimum length uncertainty principle for the physics of the
early inflationary period. Before dealing with the expanding universe
in section 3, we describe the case of flat space-time in this section.
The aim is to write down a field theory based on the modified
uncertainty (\ref{guc}). This theory is expected to have the correct
structure to describe Planck scale phenomena. As such, the mathematics
of the problem does not suggest a unique way of incorporating
(\ref{guc}) in field theory and one has to specify a prescription for
doing so. One such prescription is used in some of the earlier work on
the subject \cite{Kempf:1994su},\cite{Kempf:1996nk}. Here, we will
follow a different prescription which is suggested by, and is
self-consistent with, the physics of the problem.

\subsection{The dynamical nature of the modification}

In ordinary quantum mechanics, the Heisenberg uncertainty principle
$\Delta x \Delta p\geq \frac{1}{2}$ is a consequence of the
commutation relation $[\bx^i\,,\,\bp^j]=i\,\delta^{ij}$. This is
purely kinematic and is independent of the dynamics of the theory. 
The derivation of the uncertainty principle involves the construction
of the space of states on which the operators $\bx^i$ and $\bp^j$ act.   

It is also possible to obtain the minimum length uncertainty principle
from a modified $[\bx^i\,,\,\bp^j]$ commutation relation. In fact, in
the next section we will use such a modified commutator, following
\cite{Kempf:1994su},\cite{Kempf:1996nk}, which leads to (\ref{guc}) in
a low-energy expansion. However, a subtlety arises in this case. The
space of states on which the operators $\bx^i$ and $\bp^j$ (now
satisfying the modified commutators) act could have an involved
structure, rendering the necessary manipulations difficult. Such
representations were constructed in
\cite{Kempf:1994su},\cite{Kempf:1996nk} to which the reader can refer
for details. Fortunately, as we argue below, unlike in ordinary quantum
mechanics, now the actual details of the representation are not
relevant.

The appearance of the scale $l_s$ in (\ref{guc}) indicates that the
modification to the standard Heisenberg uncertainty principle is
dynamical in nature. In other words, if we have a fundamental theory
capable of describing physics at the Planck scale, then the
modification arises as a consequence of the dynamics and of the
particular structure of the fundamental theory; without ever having to
invoke modified $\bx^i,\bp^j$ commutators \footnote{For example,
this is how (\ref{guc}) is inferred in \cite{Gross:1987ar},
\cite{Amati:1988tn} from the behaviour of the string scattering
cross-sections at high energies. The string theory itself is based on
standard commutators.}. Thus, although one may invent $\bx^i,\bp^j$
commutators that result in (\ref{guc}), in practice, the modifications
arise totally independent of the existence of such commutators and
hence, of the structure of the function spaces on which they are
realized. 

Suppose we could derive an effective field theory description of
some high-energy process from our fundamental theory, for example, by
summing up, to all orders, the relevant terms of a perturbative
expansion in powers of energy. Such an effective field theory would
automatically incorporate the minimum length feature of (\ref{guc}) in
its structure. Unfortunately, at present we do not know how to carry
out such a computation in practice, especially in the case of interest
which is a field theory in an expanding universe. Now the notion of a
modified commutator comes handy. Instead of adding the high-energy
corrections that encode (\ref{guc}) to the low-energy effective field
theory, we regard them as getting absorbed in a modification of the
$\bx^i,\bp^j$ commutators. This will leave the apparent structure of
the field theory unchanged while modifying the relation between
coordinates and momenta. Whatever the details, this should finally
result in modified commutators consistent with the minimum length
principle. We then transcribe this effect back from the commutators to
the field theory, enabling us to surmise its high-energy modification.
It is evident that in this approach the modified commutator has no
fundamental significance. It is only used as a trick to guess the form
of the high-energy corrections to a field theory consistent with the
minimal length principle. This procedure is implemented explicitly in
sections 2.3 and 4.1.

\subsection{Modified commutation relations in flat space}

The uncertainty relation (\ref{guc}) can also be obtained from
modified $\bx^i,\bp^j$ commutators. Here we consider a general class of
modified commutation relations, consistent with rotational invariance,
that lead to a minimum length uncertainty\footnote{Throughout this
paper, bold-face letters $\bx^i, \bp^i,\brho^i,\cdots$ denote quantum
mechanical operators corresponding to $x^i, p^i,\rho^i,\cdots$.
Symbols $p, \rho, \cdots$ denote the magnitudes of vectors $p^i,
\rho^i$, {\it etc}.} \cite{Kempf:1994su},\cite{Kempf:1996nk}, 
\beq
[\bx^i\,,\,\bp^j]=i\,\delta^{ij}f(\bp)+i\,g(\bp)\,\bp^i \bp^j\,,
\qquad {[\bx^i\,,\,\bx^j]} =0\,,\qquad [\bp^i\,,\,\bp^j]=0\,.
\label{mcf}
\eeq
The information about the modification is entirely contained in the
function $f(p)$. The term with $g(p)$ is required by the Jacobi
identities and is fully determined in terms of $f(p)$.  The obvious
restriction on $f(p)$ is that for small enough $p$ it should reduce to
$1$, leading back to the standard commutators. It is understood that
$p$ is measured in units of some high-energy scale, say, the Planck
mass, $l_s^{-1}$.

To elucidate the implications of the modified commutator, we exploit a
useful construction introduced in \cite{Kempf:1996nk}: Introduce
auxiliary variables $\rho^i$ such that 
on functions $\phi(\rho)$, 
\beq
\bx^i\phi(\rho) = i\,\frac{\p}{\p\rho^i}\,\phi(\rho)\,.
\label{xrho}
\eeq
The $\rho^i$ are given in terms of the momenta $p^i$ as, 
\beq
\rho^i = \frac{p^i}{f(p)}\,.
\label{prho}
\eeq
Now, using a formal power series expansion for $f(p)$, it is easy to 
verify that (\ref{mcf}) and (\ref{xrho}) are equivalent. Further, let
us assume that $f(p)$ is such that as $p$ varies from $0$ to
$\infty$, $\rho$ stays bounded between $0$ and some maximum value 
$\rho_{max}$. Here, $p$ and $\rho$ denote the magnitudes of $p^i$
and $\rho^i$, respectively. 

One can now understand the essential consequence of (\ref{mcf}) using
a simple representation (one corresponding to a particle of momentum
``$x$'' in a rigid box of size ``$2\rho_{max}$''), without getting
into the details of the construction of the more complicated function
spaces. Equation (\ref{xrho}) implies the commutator 
\beq
[\,\bx^i, \brho^{\,j}\,]=i\,\delta^{ij}
\label{xrhof}\,,
\eeq
and hence, the associated uncertainty relation $\Delta x^i \geq
1/(2\Delta \rho^i)$. Since we assume $\rho$ to be bounded by
$\rho_{max}$, the maximum uncertainty in $\rho^i$ is $2\rho_{max}$.
Thus, there is an associated minimum length uncertainty,
\beq
(\Delta x)_{min}\sim \frac{1}{\rho_{max}}\,.
\label{xmin}
\eeq  
This is required to be of order one, in units of the $l_s$ appearing
in (\ref{guc}). The conditions on $f(p)$ are summarized below: 
\begin{itemize}
\item 
As $p\rightarrow 0$, $f(p)\rightarrow 1$ and $\rho\rightarrow p$ \,.
\item 
$\rho=p/f(p)$ is bounded by $1$ in units of $l_s^{-1}$, which we take
to be the Planck mass. 
\item
The most natural functions $f$ satisfying the last condition 
are those for which $\rho$ increases monotonically with $p$,
approaching its maximum value $\rho_{max}$ as $p\rightarrow\infty$. 
An example is 
\beq
\rho = \,\tanh^{{1}/{\gamma}}\,(p^\gamma)\,.
\label{unruh2}
\eeq
\end{itemize}
We close this subsection with two comments. In formulating a theory
based on the modified commutator, it may be tempting to speculate on
identifying $\rho$ as the physical momentum with a cut-off, as in
\cite{Kempf:2000ac}. However, a momentum cut-off is not natural from
the point of view of fundamental physics. For example, in string
theory loop diagrams are naturally regulated as a consequence of
modular invariance, and not by a momentum cut-off. Moreover, if the
modified commutator (\ref{mcf}) is obtained by analyzing physical
processes in some fundamental theory, say, through a calculation
leading to equation (\ref{mdisf}) below, then $p$ will manifestly
correspond to the physical momentum.

Furthermore, as in ordinary quantum mechanics, one faces ordering
ambiguities when dealing with products of $x^i$ and $p^j$.  Now, the
problem is more severe since the ambiguity is itself $p$-dependent. It
is therefore safest to avoid coordinate systems in which the
Lagrangian involves such products.

\subsection{Scalar fields in flat space with modified commutators}

A field theory based on the modified commutation relation (\ref{mcf})
will contain new effects that become relevant at large momenta. As
explained in subsection $2.1$, this is expected to reproduce an
effective field theory that provides a description of high energy
interactions in a fundamental theory. In this subsection we will
consider the implications for the scalar field propagator in flat
space-time. In particular we will discuss two effects that will be
relevant to the trans-Planckian problem of inflationary cosmology,
{\it i}) the modification of dispersion relation and the associated
minimum bound on the wavelengths, and {\it ii}) the increase in phase
space volume occupied by a quantum state at high momenta. The
generalization to field theory in the Robertson-Walker universe will
be considered in the coming sections.

Consider a massless scalar field $\phi(t,x)$ with the standard action
$S$, and equation of motion (ignoring interaction terms), 
\beq
S=-\frac{1}{2}\int d^4x\,\p^\mu\phi\,\p_\mu\phi\,\,,\qquad\quad \left(
\p_t^2 - \p^i\p_i \right) \,\phi(t,x)=0\,.
\label{phi-x}
\eeq 
The justification for choosing this as the starting point was given in
subsection $2.1$: Imagine that the field theory is derived from some
fundamental theory. Then, for low-energy processes one obtains
(\ref{phi-x}) along with interaction terms not considered here. To
describe very high energy processes, the field theory is expected to
get modified such that its dynamics leads to the minimum length
uncertainty (\ref{guc}). Since such modifications have not yet been
derived from first principle, we assume, alternatively, that the high
energy corrections coming from the fundamental theory can also be
accommodated by modifying the ordinary $\bx^i,\bp^j$ commutators to
the form (\ref{mcf}), keeping the action unchanged. In this case,
the action preserves its low-energy form (\ref{phi-x}), although
now the relation between ``coordinate'' and ``momentum'' has changed
and they are no longer conjugate variables, {\it i.e.}, $-i\p/\p x$ does
not represent $p$.

In fact, the discussion in the previous subsection shows that the
coordinates $x^i$ are now conjugate to the variables $\rho_i$. Thus,
we interpret the symbol ``$\p/\p x^i$'' in (\ref{phi-x}) simply as a
representation of $i\rho_i$ on the appropriate functions $\phi(x)$,
irrespective of the details of the construction (for a discussion of
the representations see \cite{Kempf:1994su,Kempf:1996nk}). In this
sense, the commutation relation (\ref{xrhof}) allows us to make a
Fourier transformation to a variable $\tphi(t,\rho)$ through
$\phi(t,x)=N\, \int^{\rho_{max}}_{-\rho_{max}}d^3\rho\,\tphi(t,\rho)\,
e^{ix^i\,\rho_i}$, where $N$ is a normalization constant. Then, the
free massless equation of motion becomes
\beq
\Big(\,\p_t^2 + \rho^2(p)\,\Big)\,\tphi(t,\rho) \,\equiv\,
\Big(\,\p_t^2 + \frac{p^2}{f^2(p)}\,\Big)\,\tphi(t,p)=0\,,
\label{mdisf}
\eeq
where, $\rho$ has been expressed in terms of $p$ using (\ref{prho}).
We have used the same notation $\tilde\phi$ for the field as a
function of both $\rho$ and $p$; the difference should be clear from
the context. 

To further clarify the origin of the modification, let us introduce 
new variables $\hat{\bf x}^i$ such that their commutators with $\bp^i$
have the standard form, 
\beq
[\,\hat{\bf x}^i, \bp^{\,j}\,]=i\,\delta^{ij}\,.
\label{xhatp}
\eeq
In the $\hat x$-representation, $p_i=-i\p/\p{\hat x}^i$. Then, in
terms of $\hat x$ the modified equation of motion is equivalent to the
higher-derivative equation, 
\beq
\Big[\,\p_t^2 + \rho^2(-i\p_{\hat x})\,\Big]\hat\phi(t,\hat{x})=0\,. 
\label{hatphi}
\eeq
The new variables $\hat x^i$ are identified as the ordinary space
coordinates. 

One may take the approach that (\ref{mdisf}) is the basic equation
that emerges in the {\it effective field theory limit} of some
fundamental theory in a momentum-space computation, in the spirit of
\cite{Kaloper:2002uj}. This could then be given a space-time
representation either in terms of the two-derivative field equation
(\ref{phi-x}) with modified commutators, or in terms of the
infinite-derivative field equation (\ref{hatphi}) with the standard
commutators. In this sense, the use of the modified commutators
enables us to express the higher derivative corrections to the lowest
order equation of motion in a closed form, consistent with the notion
of a minimal length scale. The closed form of the expression is
crucial since a truncated power series expansion is not a good
approximation at large momenta. In the string theory context, this is
analogous to summing up the $\alpha^\prime$ corrections to all orders
(albeit for the issue of Lorentz invariance on which we will comment
below). To summarize, at low energies one may start with (\ref{phi-x})
and the standard commutation relations. Then high energy corrections,
assuming that they are summable, will convert this to equation
(\ref{hatphi}), keeping the commutators unchanged. Alternatively, one
may incorporate the effect of high energy corrections into the
commutation relation keeping the form of the action unchanged.

The interpretation of (\ref{mdisf}) as the basic equation will turn
out to be particularly useful when we consider the Robertson-Walker
space-time later. Then the analogue of this equation is still well
defined even though it no longer has a space-time representation
analogous to equation (\ref{phi-x}).

As indicated above, Lorentz invariance is broken by the modified
commutator. As a result, the time evolution of the solutions is still
determined by a second order equation, in spite of the higher space
derivatives. For comments on the possibility of a dynamical breakdown
of Lorentz invariance near the Planck scale, see
\cite{Jacobson:1991gr}. One may also circumvent the issue of Lorentz
invariance and regard the flat space field theory of this section only
as a toy example employed to illustrate the ideas. Our aim is to
finally use the modified commutators in Robertson-Walker space-times,
where Lorentz invariance is anyway broken by the background. For a
related setup, where a time-dependent background leads to a modified
dispersion relation, see \cite{Amelino-Camelia:1996pj}.

Independently of its relevance to cosmology, equation (\ref{mdisf})
defines an interesting field theory in its own right.  The built-in
minimum length uncertainty in the field theory may also render it more
finite as can be inferred from the finite range of $\rho$-integration,
although the physical momentum remains unbounded.

Equation (\ref{mdisf}) contains the non-linear dispersion relation
$\omega=\rho(p)$ which is fully determined by the function $f(p)$ and
satisfies the restrictions discussed above. For example, for the
choice (\ref{unruh2}) one gets the Unruh dispersion relation which has
been used to mimic the effects of trans-Planckian physics in the
contexts of Hawking radiation \cite{Unruh:1995} and in inflationary
cosmology \cite{Martin:2000xs}. Let us now consider the minimum
wavelength bound implied by the modified dispersion relation. From
(\ref{phi-x}) and (\ref{xrhof}) it is obvious that the plane wave in
the theory with the modified commutator is given by $e^{i\omega
t+i\rho_ix^i}$. This has a wavelength,
\beq
\lambda =\frac{2\pi}{\rho} = {2\pi}\, \frac{f(p)}{p}\,.
\label{wlf}
\eeq
Since $\rho$ is bounded by $1/l_s$, the wavelength of a fluctuation
is never smaller than the Planck length regardless of how high the
value of its momentum may be. Also, the energy $\omega =\rho$ is
always bounded by the Planck mass.  

When the Fourier transform is implemented in the action $S$, then the
change of variable from $\rho$ to $p$ also induces a Jacobian which
can be computed as $J=\p(\rho^3)/\p(p^3)$,
\beq
\ds S=\frac{1}{2}\int dt\,d^3 p\,J(p)\, 
\Big(\p_t\tphi^*\,\p_t\tphi - \rho^2(p)\,\tphi^*\,\tphi \Big)\,. 
\label{Sp} 
\eeq
In flat space-time, the Jacobian does not affect the equation of
motion for $\tilde\phi(t,p)$. However, since $\int d^3 p\,J(p)/
\hbar^3$ corresponds to a sum over momentum states (where $\hbar$ has
been reinstated), it indicates that now a momentum mode $p$ occupies
phase space volume $\sim \hbar^3/J$ as opposed to $\sim \hbar^3$ of
ordinary quantum mechanics. For a $\rho(p)$ consistent with the
minimum length uncertainty, $J$ decreases for large $p$ and hence the
phase space volume occupied by a momentum mode $p$ increases.
Equivalently, the total number of degrees of freedom at high energies
decreases. This is consistent with the results of \cite{Atick:1988si}
which finds a similar behaviour for the degrees of freedom in string
theory at high temperatures. This further supports our prescription
for incorporating the modified commutation relations in field theory.
As we will see later, in an expanding universe, the Jacobian affects
the equation of motion in a non-trivial way and has some interesting
consequences.

The modification of the dispersion relation and the associated bound
on the wavelength, makes the modified commutator theory very appealing
from the point of view of the trans-Planckian problem in inflationary
cosmology. In this latter context, modified dispersion relations
similar to (\ref{mdisf}) have been proposed
\cite{Martin:2000xs},\cite{Brandenberger:2000wr},
\cite{Martin:2000bv}-\cite{Lemoine:2001ar},\cite{Mersini:2001su},
albeit in an {\it ad hoc} manner (except for
\cite{Kowalski-Glikman:2000dz}, and in a related context
\cite{KalyanaRama:2001xd}), to encode the effects of trans-Planckian
physics on the dynamics of scalar fields. Furthermore, the form of
(\ref{wlf}) may be taken to indicate that wavelengths corresponding to
the present day scales and CMBR anisotropies, when blue-shifted
backward in time to the beginning of the inflationary period, will
never attain sub-Planckian lengths, regardless of how long the
inflation may last. As we will see, these features carry over to the
case of inflationary universe, although the time dependence of the
physical momenta, due to the red-shift, introduces some subtleties as
well as new effects.

\section{The modified commutator in flat RW space-time}

To apply the modified commutator to the early universe, it should
first be generalized to the flat Robertson-Walker (RW) space-time.
This generalization will be carried out in the present section. The
underlying assumption, of course, is that a geometric description of
the space-time still remains valid, at least in some approximation.
This is reasonable since the effects of the modified commutator will
be perceived by the high-energy fluctuations and not by the slowly
varying background fields.

We recall the flat RW metric in the conformal frame,
\beq
ds^2=a^2(\eta)\left(-d\eta^2 + dy^i dy^j\delta_{ij}\right)\,. 
\label{dS}
\eeq
Here, $\eta$ is the conformal time and $y^i$ are comoving coordinates.
Physical distances at time $\eta$ are simply given by $a(\eta)\,y^i$.
Let us denote the comoving momentum (wave number) conjugate to $y^i$
by $k_i$. Then, the physical momentum undergoes a red-shift with time
and, in the metric above, is given by $k_i/a(\eta)$. The minimum
length uncertainty should be time independent and apply to the
physical lengths. At the same time, the modified commutator should be
consistent with the residual symmetries of the metric (\ref{dS}), in
terms of comoving coordinates.

Only a subset of general coordinate transformations keep the form of
the RW metric in comoving coordinates unchanged. Besides rigid spatial
rotations and translations, which are in common with flat space, this
subset also contains constant rescalings of the coordinates that
amount to rescaling $a$, keeping the form of the metric unchanged.
Then, the analogue of the modified commutator (\ref{mcf}) in the flat
RW space, consistent with the residual covariance of the metric ${\rm
g_{\mu\nu}}$ in the comoving coordinates, is
\beq
[\;\by^i\;,\,\bk^j\;] = i{\rm g}^{ij}\,f(\bk) +
ig(\bk)\,\bk^i\,\bk^j\,,
\label{mcdS}
\eeq
where, $k^i={\rm g}^{ij}k_j = a^{-2}\delta^{ij}k_j$. This should not
lead to a fixed minimum length uncertainty in the comoving coordinates
$y^i$, which would result in a rather large uncertainty in the proper
distance $ay^i$ at the present epoch. It is easy to see that it is
actually the physical distance that has a fixed minimum uncertainty:
In order to compare with the flat space case, let us introduce
$n_i=k_i$, $n^i=\delta^{ij}n_j=a^2 k^i$, so that $k^2=n^2/a^2$. In
terms of this, the commutation relation takes the form
\beq
[a\by^i\,,\,\frac{\bn^j}{a}] = i\delta^{ij}\,f(\frac{\,\bn}{\,a})+
ig(\frac{\,\bn}{\,a})\,\frac{\bn^i}{a}\,\frac{\bn^j}{a}\,.
\label{mcph}
\eeq
This has the same structure as (\ref{mcf}) and hence implies a
minimum position uncertainty in the physical or proper distance 
$a y^i$ as desired, with $n^i/a$ as the physical momentum. More
precisely, in analogy with the flat space case, one can introduce the
``physical'' auxiliary variables $\rho_i$ as well as ``comoving''
ones, $\hat\rho_i$, given by 
\beq
\rho_i = \frac{n_i/a}{f(n/a)}\,, \qquad\qquad \hat\rho_i=a\,\rho_i\,.  
\label{rhon}
\eeq
The commutator (\ref{mcph}) now becomes
\beq
[a\by^i\,,\,\brho^{\,j}]=i\,\delta^{ij}
\label{ayrho}
\eeq
This leads to a minimum length in $ay^i$ as $(a\Delta y)_{min}
\sim\rho_{max}^{-1}$. Since $\rho$ is a function of a single variable,
this value is $\eta$-independent. In terms of the ``comoving'' version
of the auxiliary variable $\hat\rho_i= a\rho_i$, the commutator is
$[\by^i\,,\,\hat{\brho}_{j}] =i\,\delta^i_{\,j}$. Note that unlike the
flat space case of the previous section, now $f$ and $\rho_i$ have
become time dependent through the red-shift of the physical momentum.
However, they still satisfy the restrictions described below equation
(\ref{xmin}).

\section{Trans-Planckian effects in the inflationary era}

Of late, it has been customary to assume that the effects of
fundamental physics at Planckian energies can be effectively
incorporated in field theory by modifying the free field dispersion
relation in some appropriate way
\cite{Martin:2000xs}-\cite{Shankaranarayanan:2002ax}. In practice, in
the absence of a first principle approach, one ends up choosing the
modified dispersion relations in an {\it ad hoc} manner. This approach
has been used to investigate the possible effects of the Planck scale
physics on the CMBR spectrum during the early inflationary epoch. In
this section we attempt to achieve a more fundamental understanding of
this issue based on the modified uncertainty principle in RW universe.
It will be shown that the time-dependent non-linear dispersion
relation emerges in a natural way and its form is determined by the
modified commutator. There is also a further modification associated
with the reduction in the number of high energy degrees of freedom.
This could dominate at very high momenta with the effect of dampening
the fluctuations. This new feature will be investigated in detail in
section 5.

\subsection{Fields in flat RW space with modified commutators}

We will now consider the implications of the modified commutation
relations for scalar field theory in flat Robertson-Walker space-time.
The resulting theory includes the effects of Planck scale physics in
an inflationary universe. The momentum red-shift associated with the
rapid expansion induces new effects beyond those considered in section
2. Let us start with the standard massless scalar field theory in
Robertson-Walker space-time, 
\beq
S=-\frac{1}{2}\int d^4x\sqrt{-{\rm g}}\,{\rm g}^{\mu\nu}
\p_{\mu}\phi\,\p_{\nu}\phi =
\frac{1}{2}\int d\eta\,d^3y\,a^2\Big(\p_{\eta}\phi\,\p_{\eta}\phi
-\sum_{i=1}^3 \p_{y^i}\phi\,\p_{y^i}\phi\Big)\,. 
\label{SdS}
\eeq
Here, $y$ is regarded as a variable conjugate to $a\rho$ and not to the
comoving momentum $n$. Of course, it is not obvious that such a simple
looking space-time description could survive near the Planck scale (we
will see indications that it does not). But in the absence of a better
formalism, we choose this as our starting point, in analogy with the
flat space case considered in section 2. The picture described there
is still valid with some refinements to be discussed.

Since $\by^i$ and $\hat\brho_i$ satisfy the standard commutation
relation, we simply interpret the symbol ``$-i\p_{y^i}$'' in 
 (\ref{SdS}) as the operator conjugate to $\hat\brho_i=a\brho_i$. In
order to rewrite the action in the $\rho$-representation, consider the
Fourier transform 
\beq
\phi(\eta,y)= a^3 \,\int^{l_s^{-1}}_{-l_s^{-1}}d^3 \rho
\,\tphi(\eta,\rho)\, e^{iy^i a \rho_i}\,,
\label{ft}
\eeq 
which in the limit $l_s\rightarrow 0$, $\rho\rightarrow n/a$ reduces to
the standard expression in the absence of Planck scale effects. Then
by the above interpretation,
\beq
-i\mbox{``$\p_{y^i}$''}\,\phi(\eta,y)=
a^3 \,\int^{l_s^{-1}}_{-l_s^{-1}} d^3 \rho
\,\tphi(\eta,\rho) a\rho_i \, e^{iy^i a \rho_i}\,,
\label{dyphi}
\eeq 
Also, since $\rho$ is a dummy integration variable, 
\beq
\p_\eta\phi= a^3 \,\int^{l_s^{-1}}_{-l_s^{-1}}d^3 \rho\, 
(\,\p_\eta\tphi\,+ 
\Bigg[\,i\,y^ia\rho_i+3\Bigg]\frac{\p_\eta a}{a}\tphi\,)
\,e^{iy^i a\rho_i}\,.
\label{detaphi}
\eeq
The correct $l_s\rightarrow 0$, $\rho\rightarrow n/a$ limit of this
is reproduced by the term outside the square brackets alone. In
this sense, the two terms within the square brackets conspire to
cancel in this limit. To see this, note that both terms in the square
brackets are needed to assemble the right hand side as the
$\eta$-derivative of the right hand side of (\ref{ft}), where the limit
can be easily taken. For example, if $3\p_\eta a/a\tphi$ is dropped, 
it will no longer be possible to recover the correct limit from 
(\ref{detaphi}).

To make sense of carrying out the Fourier transform in (\ref{SdS}),
using the above expressions, we choose the simple representation in
which $y^i a$ is discrete and interpret $\int d^3y$ in (\ref{SdS}) as
denoting a sum (as argued earlier the final result is independent of
the representation used). Then at any given time $\eta$,
\beq
y^i a\rightarrow \pi m^i l_s\,,\qquad
\int d^3y \rightarrow \frac{1}{a^3}\left(\frac{l_s}{2}\right)^3
 \prod_{i=1}^{3}
\sum_{m^i=-\infty}^{+\infty}\,.
\label{disc}
\eeq
We can now use (\ref{ft})-(\ref{disc}) to write the action in the
$\rho$-representation. Let us first ignore the terms within the
square brackets in (\ref{detaphi}) (we will argue below that the action
cannot contain contributions of this form). Then, after expressing
$\rho$ in terms of $n$ from (\ref{rhon}), the $\rho$-space action
yields, 
\beq
\ds S=\frac{1}{2}\int d\eta\,d^3 n\,J(\frac{n}{a})\, a^2
\Big(\p_{\eta}\tphi^*\,\p_{\eta}\tphi - \frac{n^2}{f^2(n/a)}\, 
\tphi^*\,\tphi \Big)\,. 
\label{Sn} 
\eeq
We use the same notation $\tilde\phi$ for the field as a function of
both $\rho$ and $n$ since the difference is clear form the
context. Here, $n^2=\sum_{i=1}^3 n_in_i$ and $J(n/a)$ is the Jacobian
for the change of variables determined by (\ref{rhon}),  
\beq
J(v)=\frac{\rho^2}{v^2}\,\frac{\p \rho}{\p v}\,,\qquad\quad
\left(v=\frac{n}{a}\right)\,.
\label{J}
\eeq
The restrictions on $\rho$ fix the asymptotic behaviour of $J$ such
that $J\rightarrow 1$ for $v\ll 1$, and $J\rightarrow 0$ for large
physical momentum $v$. The equation of motion following from
(\ref{Sn}) is 
\beq
\ds\left(\,\p_\eta^2\, +\, a^2\,\rho^2(\frac{n}{a})
\,-\,\frac{\p_\eta^2 ({\sqrt J}\,a)}{{\sqrt J}\,a}\,\right)
\tilde\chi(\eta,n)=0\,,
\label{eomn}
\eeq
where $\tilde\chi(\eta,n)=a\,{\sqrt J}\,\tilde\phi(\eta, n)$. This is  
the generalization of equation (\ref{mdisf}) from flat space-time to the
RW space-time. 

Before proceeding further, let us get back to the terms within the
square brackets in (\ref{detaphi}) that have been ignored. The first
term in the square brackets will appear in (\ref{Sn}) and (\ref{eomn})
as $\rho_i\p_{\rho_i}\tphi$. It is not difficult to see that such
expressions should not appear at all: We have been regarding
our modified theory as an effective field theory description of
high energy processes in some fundamental theory. An explicit
derivation of the effective theory from the fundamental theory (if it
could be carried out) would then involve momentum space computations,
as is generically the case in such derivations (see
\cite{Kaloper:2002uj} for a related approach). For a spatially flat
metric, such computations cannot lead to expressions containing
momentum space derivatives, $\p_p\tphi(p)$, which are required to
produce the terms we have ignored. In the coordinate representation,
such expressions would lead to space-dependent potentials which are
not consistent with the homogeneity of the background and hence are
ruled out by it.

If for the above reason we ignore the first term in the square brackets
in (\ref{detaphi}) but retain the second term, then our modified
theory will not reduce to the standard theory of scalar fields in RW
space-time in the limit $l_s\rightarrow 0$, when Planck scale effects
disappear. This is evident from the discussion below (\ref{detaphi}).
Since this is not desirable, we conclude that both terms within the
square brackets in (\ref{detaphi}) should be ignored.

To summarize, we have argued that scalar field theory in flat RW
space-time, based on the modified commutators (\ref{mcdS}), or
(\ref{mcph}), is directly defined by (\ref{Sn}) and (\ref{eomn}),
without the extra terms. This also was our preferred interpretation of
the analogous equation (\ref{mdisf}) in flat space. However, while
there the associated coordinate space equations could be written in
the usual form (in terms of a coordinate conjugate to $\rho$), in the
present example of an expanding universe this is not the case. Thus,
the space-time action (\ref{SdS}) is a useful starting point, although
it is incomplete.

There is a further conceptual difference between (\ref{mdisf}) and
(\ref{eomn}): The modified uncertainty relation in flat space-time,
which leads to (\ref{mdisf}), can be regarded as the outcome of a
string theory computation. As for (\ref{eomn}), a similar statement
could in principle be made, although in practice string theory in the
expanding universe is still poorly understood. For a recent discussion
of the difficulties see \cite{Balasubramanian:2002ry}.

\subsection{Time-dependent non-linear dispersion relations}

The standard equation of motion for a scalar field in RW metric
corresponds to the small momentum limit of (\ref{eomn}), where
$a\rho\sim n$ and $J\sim 1$. The effects of Planck scale physics, as
encoded in the modified commutation relations, result in the
modification of the dispersion relation and in the appearance of $J$
in the equation of motion. The $J$ dependence is a novel feature and
its consequence will be investigated in the next section. Here we
discuss the dispersion relation.

In the equation of motion (\ref{eomn}), the free field dispersion
relation $\omega^{free}_{phys}=n/a$, is modified to a non-linear, time
dependent one,    
\beq 
\omega_{phys} = \rho \equiv \frac{n/a}{f(n/a)}\,.  
\label{mdis}
\eeq
This is a generalization of the modified dispersion relations
considered in \cite{Martin:2000xs},\cite{Brandenberger:2000wr},
\cite{Martin:2000bv}-\cite{Shankaranarayanan:2002ax}. However, the
modification is no longer {\it ad hoc} (as is also the case in
\cite{Kowalski-Glikman:2000dz}), but is determined by the nature of
the uncertainty principle (\ref{mcdS}), through the function $f$. As
described in subsection 2.2 (following equation (\ref{xmin})), this
function satisfies certain restrictions. In terms of
$\omega_{phys}^2$, the restrictions translate to the following:
\begin{itemize}
\item The linear dispersion relation $\omega_{phys}=n/a$ emerges at
small physical momenta $n/a$.
\item $\omega_{phys}^2\geq 0$ and $\omega_{phys}$ is always real.
\item The quantity $\omega_{phys}$ is bounded by $\rho_{max}$ which
is fixed by the minimum length scale as $1/l_s$.
\end{itemize}
Thus, while there is no cut-off on the physical momentum $n_i/a$,
there is always a cut-off on the effective frequency $\omega_{phys}$.
These restrictions exclude some of the dispersion relations that have
appeared in the literature as being inconsistent with the minimum
length uncertainty. Some often used dispersion relations are
depicted in Figure 1.    

The scalar field theory with the modified dispersion relation can now
be second quantized in terms of $\tilde\chi(\eta, n)$, in the usual
fashion. The modification of the commutation relation in the first
quantized theory does not affect the field commutators of the second
quantized theory.

As stated in section 2, the most natural class of solutions to the
restrictions on $f$ consist of $\rho$ increasing monotonically with
$n/a$. This avoids the problem of associating multiple momenta with
the same frequency. For example, the class of solutions in
(\ref{unruh2}), lead to  
\beq
\omega_{phys} = \,\tanh^{1/\gamma}\,\left(\frac{n}{a}\right)^\gamma\,. 
\label{unruh-gamma}
\eeq
These correspond to the well known Unruh dispersion relations
\cite{Unruh:1995}. For these, equation (\ref{eomn}) in the special
case of $J=1$, was analyzed in \cite{Martin:2000xs}. The conclusion
was that this class of trans-Planckian modifications does not modify
the spectrum of cosmological perturbations calculated with the linear
dispersion relation. The case of $J\neq 1$ will be discussed in some
detail in the next section.

Another form of $f$ appearing in the literature is \cite{Kempf:1996nk} 
\beq
\ds f(n/a)= \frac{2 (n^2/a^2)}{\sqrt{1+4(n^2/a^2)}\,-1}\,.
\label{soft}
\eeq
\begin{wrapfigure}{r}{9cm}
\epsfxsize=9cm
\centerline{\epsffile{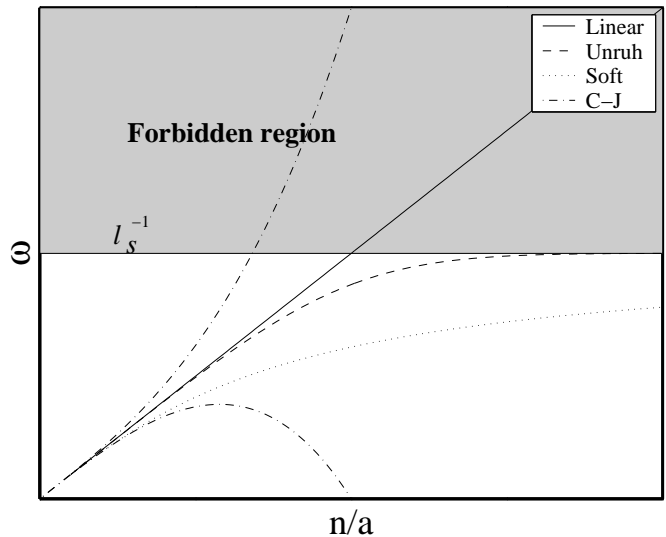}}    
\caption{\small Dispersion relations}
\end{wrapfigure}
\par\noindent
One can numerically verify that the dispersion relation corresponding
to this function has the same qualitative behaviour as the Unruh form
(\ref{unruh-gamma}), with an appropriate value of $\gamma$. In Figure
1, this curve is labeled as ``soft''. Hence one can infer that for
$J=1$, it leaves the spectrum of cosmological perturbations unchanged
(later we comment on $J\neq 1$). This conclusion is at variance with
the result of \cite{Kempf:2000ac} (and the follow-up analysis in
\cite{Kempf:2001fa}, \cite{Easther:2001fz})
which uses the same form of $f$, but implements the modified
commutator in a different manner.
\par\par
Next, we consider the generalized Corley-Jacobson type of dispersion
relations \cite{Corley:1996ar},\cite{Martin:2000xs},
\beq
\rho^2 =\frac{n^2}{a^2}\left(1+\sum_{q=1}^m b_q\,
(\frac{\,n^2}{\,a^2})^{q}\right)\,.
\eeq
Since $\rho$ is not bounded, such dispersion relations cannot be 
associated with a modified uncertainty principle with a minimum length 
scale. However, one could regard these as low-momentum expansions of
expressions that are bounded. Then $\rho$ will remain bounded as long
as the expansion is valid. Contrary to the previous two cases, the
Corley-Jacobson dispersion relations need not be monotonic functions 
of $n$. For $J=1$ the effect of these on the cosmological perturbation
spectrum has been studied in
\cite{Martin:2000xs},\cite{Brandenberger:2000wr},\cite{Martin:2000bv} - 
\cite{Brandenberger:2002hs},\cite{Niemeyer:2001qe} where it was found
that they could give rise to deviations from the predictions of linear
dispersion relation when $b_m<0$.

\subsection{Modified red-shift and evolution of wavelengths}

As pointed out, for example, in
\cite{Brandenberger:1999sw},\cite{Martin:2000xs}, in models where
inflation lasts for a sufficiently long period of time, the CMBR
anisotropies at the present epoch seem to have their origin in
fluctuations whose wavelengths, in the beginning of inflation, were
smaller than a Planck length. This assumes the validity of the
standard red-shift formula for the physical wavelength,
$\lambda_{phys}=2\pi\,a(\eta)/n$, down to the Planck length and
beyond. One may try to understand, at least heuristically, the time
evolution of wavelengths, based on the modified commutators. As
described in section 2.3, in flat space, the modified commutators lead
to a wavelength (\ref{wlf}) that cannot decrease below the Planck
length. One may extrapolate the same result to the flat RW case, in an
adiabatic approximation, simply by replacing the flat space momentum
$p$ by the physical momentum in the expanding universe, $n/a$,
\beq 
\ds\lambda =\frac{2\pi}{\rho} = 2\pi\,\frac{a}{n}\,f(\frac{n}{a})\,.
\label{wlch}
\eeq
\begin{wrapfigure}{r}{8.5cm}
\epsfxsize=8.5cm
\centerline{\epsffile{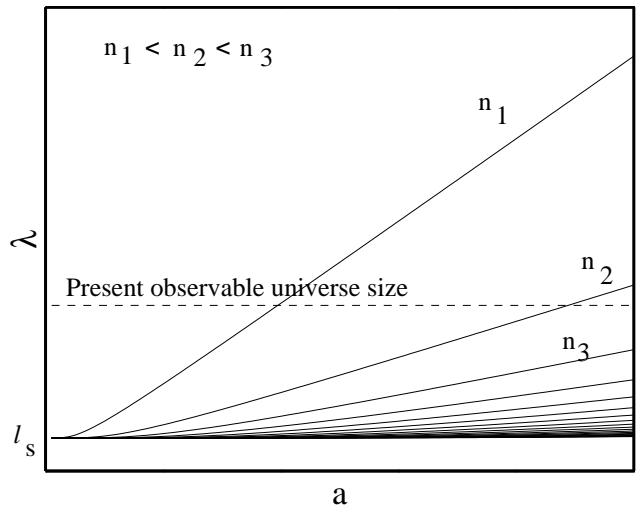}}    
\caption{\small Time evolution of wavelength $\lambda$ associated with
momentum mode $n$} 
\end{wrapfigure}
We take $\rho$ to be a monotonically increasing function. At late
times (large $a$), the function $f$ tends to identity and one recovers
the usual red-shift formula. As we go backward in time (decreasing
$a$), $\lambda$ tends to a constant $2\pi$ (in units of $l_s$). Near
$a=0$, a small wavelength band corresponds to a large momentum
interval. This behaviour is shown in Fig. 2 for an $f$ associated with
the Unruh dispersion relation (\ref{unruh-gamma}) with $\gamma=1$.
Thus equation (\ref{wlch}) insures that at the end of the inflationary
era one still finds fluctuations of all possible wavelengths within
the observable universe without requiring sub-Planckian wavelengths in
the early epochs. This argument, although reasonable, still remains
heuristic since a plane wave with the above wavelength is not really a
solution of the equation of motion (\ref{eomn}), due to the time
dependence of $a$. However, since the theory has an inbuilt minimum
length uncertainty, this picture does capture the qualitative
behaviour of length scales associated with the fluctuations.

\section{Trans-Planckian damping and the initial state}

In this section we discuss the consequences of the appearance of $J$
in the last term of equation (\ref{eomn}). To put our results in
context, we first briefly review the relevant features of the $J=1$
case analyzed in \cite{Martin:2000xs}. It is then argued that the
presence of $J$ can help us fix the initial state of the scalar 
field. 

\subsection{Review of the $J=1$ case}

When the dispersion relation is modified by hand, one misses the
Jacobian $J$ in the equations. This corresponds to $J=1$ for which
equation (\ref{eomn}) reduces to   
\beq
\ds\left(\,\p_\eta^2\, +\, a^2\,\rho^2(v)
\,-\,\frac{\p_\eta^2 a}{a}\right)
\tilde\chi_{MB}(\eta,n)=0\,.
\label{eomJ1}
\eeq
This is the equation that has so far been used to study the
cosmological implications of modified dispersion relations. It was
analyzed in detail by Martin and Brandenberger \cite{Martin:2000xs}
for the Unruh and Corley-Jacobson type dispersion relations. Here we
review some features of this model which are of interest to us. We
assume de Sitter expansion for which $\p_\eta^2 a/a=2 H^2 a^2$. Let
us follow a mode of comoving momentum $n$ backward in time, as its
physical momentum, $v=n/a$, is blue-shifted. The momentum range can be  
divided into three regions according to the behaviour of the solution: 
\begin{itemize}
\item At late times, when $v<<H$ (region III), the dispersion relation
is linear and $\rho^2=v^2$ can be neglected as compared to $\p_\eta^2
a/a$. The relevant solution is $\tilde\chi^{\rm (III)}_{MB}=C^{\rm
(III)}_n\,a$.
\item When $H<<v<<1$ (region II), the dispersion relation is
still linear, but now $\p_\eta^2 a/a$ can be neglected and one gets an
oscillatory solution $\tilde\chi^{\rm (II)}_{MB}=C^{\rm
(II)}_{1n}\,e^{in\eta}+C^{\rm (II)}_{2n}\,e^{-in\eta}$. If the
dispersion relation had not been modified, then region II would have
extended beyond the Planck scale, $v=1$, all the way up to the
beginning of inflation. This is the standard scenario that consequently
suffers from the trans-Planckian problem.
\item For non-linear dispersion relations, when $v>1$ one enters
region I. Then, for the Unruh dispersion relation, $\rho^2\simeq 1$.
In the de Sitter phase, $a^2=(H\eta)^{-2}$ and the equation has an
exact solution, $\tilde\chi^{\rm (I)}_{MB}= C^{\rm (I)}_{1n}
|\eta|^{\delta_1}+ C^{\rm (I)}_{2n} |\eta|^{\delta_2}$. The exponents
are determined in terms of $H$, $\delta_{1,2}=\frac{1}{2}\pm
\frac{1}{2}\sqrt{9-4H^{-2}}$, and contain imaginary components.
Therefore, the solution in region I has an oscillatory nature. In the
case of Corley-Jacobson dispersion with $b_{m}<0$, at early enough
times, $\rho^2$ becomes negative and the solution is damped.
\end{itemize}
The power spectrum of scalar fluctuations is given by $\mathcal{P}=n^3
|C^{\rm (III)}_n|^2/2\pi^2$. The coefficient $C^{\rm (III)}_n$ is
determined in terms of $C^{\rm (II)}_{1n,2n}$ and $C^{\rm
(I)}_{1n,2n}$ by matching the solution and its first derivative across
the boundaries of regions III, II and I. The coefficients $C^{\rm
(I)}_{1n,2n}$ are in turn determined by the initial conditions at some
(arbitrary) time $\eta_i$ in region I, which extends to the beginning
of the inflationary era. Depending on the dispersion relation, the
power spectrum may or may not acquire a non-trivial dependence on the
initial state in the form of a modified dependence on $n$. As such,
there is no natural choice for the initial state and, for oscillatory
solutions, the best one can do is to pick up the local vacuum state at
time $\eta_i$. This is based on the implicit assumption that the modes
are created in their ground state and that nothing drastic happens
from the beginning of inflation until time $\eta_i$. Below we
consider a model based on the the equation of motion with $J\neq 1$
where the initial state problem can be addressed.

\subsection{$J\neq 1$ and trans-Planckian damping} 

In this subsection we will study the equation of motion with the
Jacobian factor $J$ included. This factor takes into account the
reduction in the number of degrees of freedom at high energies. As an
explicit example, we will concentrate on the Unruh dispersion relation
with $\gamma=3$, although the discussion can be generalized to other
cases. The solution is first discussed at a qualitative level. At the
end we will comment on the dispersion relation based on (\ref{soft})
which leads to a qualitatively different result. For convenience we
reproduce the relevant equation, (\ref{eomn}), below,
\beq
\left(\,\p_\eta^2\,+\omega_{total}^2\,\right)\tilde\chi(\eta,n)=0\,,
\qquad
\omega_{total}^2= a^2\,\rho^2(v)\,-\,
\frac{\p_\eta^2 (a\,{\sqrt J})}{a\,{\sqrt J}}\,.
\label{eomn-2}
\eeq
Here, $v=n/a$ is the magnitude of the physical momentum and 
$\tilde\chi(\eta,n)$ is the canonically normalized field given
by 
\beq
\tilde\chi(\eta,n)=a\,{\sqrt J}\,\tilde\phi(\eta, n)\,.
\label{chiphi}
\eeq
The second term in $\omega_{total}^2$ is
\beq
\frac{\p_\eta^2 (a\,{\sqrt J})}{a\,{\sqrt J}}=
\frac{v^2}{2}\left[\frac{\p_v^2J}{J}-\frac{1}{2}
\left(\frac{\p_v J}{J}\right)^2\right]\left(\frac{\p_\eta a}{a}\right)^2
+\left[1-\frac{v}{2}\,\frac{\p_v J}{J}\right]
\left(\frac{\p_\eta^2 a}{a}\right)\,.
\label{d2Ja}
\eeq 
The behaviour of this term depends on the choice of the dispersion
relation which determines $J$ through (\ref{J}). During de Sitter
expansion with Hubble parameter $H$ one has  
\beq
\left(\frac{\p_\eta a}{a}\right)^2=H^2\, a^2\,,\qquad
\left(\frac{\p_\eta^2 a}{a}\right)=2 H^2\, a^2 \,.
\label{dadeS}
\eeq
Therefore the $J$-dependent term is suppressed by the small number $H^2$.

For concreteness we will now consider the Unruh dispersion relation
(\ref{unruh-gamma}) for $\gamma=3$. The variations of $\rho^2(v)$ and
$a^{-2} \p_\eta^2(a\,{\sqrt J})/(a\,{\sqrt J})$ with $v$ are shown in
Figure 3. We have taken $H=10^{-5}$, in Planck units. Below the Planck
scale, $v<1$, (regions III and II) the modifications due to the
modified uncertainty relation are absent. 
So we will concentrate on
the trans-Planckian regime. As $v$ increases above the Planck scale in
region I ($1<v<32$), $\rho^2$ soon approaches $1$ and $J$ (not shown
in the figure) decreases rapidly. However, the $J$-dependent term
grows large in spite of the suppression by $H^2$. At the boundary
between regions I and 0, the two terms are equal and
$\omega_{total}^2=0$. 
\begin{wrapfigure}{r}{8.5cm}
\epsfxsize=8.5cm
\centerline{\epsffile{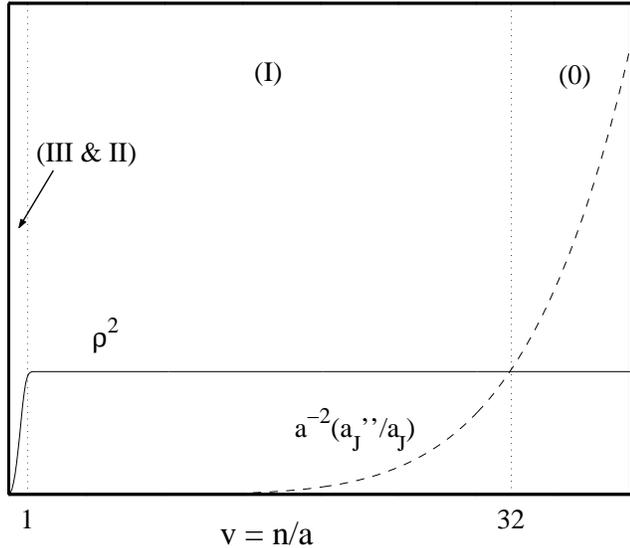}}    
\caption{\small $\rho^2$ and $a^{-2}\p_\eta^2 a_J/a_J$ 
($a_J\equiv a{\sqrt J}$) for the Unruh dispersion relation for 
$\gamma =3$, as a function of momentum $v$}   
\end{wrapfigure}
Beyond this, we enter region 0 where the
$J$-dependent term dominates and $\omega_{total}^2<0$. 
Note that the
appearance of imaginary frequencies in this model is not built by hand
into the dispersion relation. Rather, it is a consequence of the rapid
expansion during which the presence of $J$ and the red-shift
of physical momenta induce the trans-Planckian damping term. In
other words, in Planck scale processes at the present epoch, the
frequencies are real and are given by $\rho^2>0$ alone. In this sense,
the appearance of imaginary frequencies is not an unphysical feature
of the theory. In particular, one avoids the multiple valuedness of
momenta for a given energy which is a feature of non-monotonic
dispersion relations considered earlier in the literature \footnote{To
compare with the earlier models, note that although we started with
the Unruh dispersion relation, $\omega_{total}$ is reminiscent of the
Corley-Jacobson case with $b_{m}<0$.}. Region 0 is a new region that
has not appeared in earlier models.

Deep in region $0$, we can ignore $\rho^2$ in (\ref{eomn-2}).
The equation then has an obvious solution,
\beq
\tilde\chi_1^{(0)}(\eta,n) \approx  a\,{\sqrt J}\, C^{(0)}_{1n}
\sim a\, e^{-v^3}\, C^{(0)}_{1n}
\label{sol01}
\eeq
where $C^{(0)}_{1n}$ is an $n$-dependent constant. This is similar to
the solution in region III. But now the modes are frozen because, due
to the presence of ${\sqrt J}$, they perceive the universe as
expanding much faster than it really does. The other solution is 
\beq
\tilde\chi_2^{(0)}(\eta,n)=a\,{\sqrt J}\int^\eta \frac{d\eta}{a^2J}
\approx\frac{a}{{\sqrt J}}\,C^{(0)}_{2n}\sim a\,e^{v^3}\,C^{(0)}_{2n} 
\label{sol02}
\eeq
Later it will be shown that the dependence on $n$ is given by
$C^{(0)}_{1,2n}=n^{-3/2}C^{(0)}_{1,2}$. Comparing with (\ref{chiphi})
one finds that $\tilde\phi^{(0)}_1= C^{(0)}_{1n}$ and
$\tilde\phi^{(0)}_2=C^{(0)}_{2n}/J$. The energy density in terms of
the field $\tilde\phi$ is given by,
\beq
\varepsilon(v)\,d^3v\sim \frac{1}{2}\,v\,J\left(
(\p_\eta\tilde\phi)^2+a^2\rho^2\tilde\phi^2\right)d^3v
\label{edensity}
\eeq
As we go backward in time, the energy density for the first solution
vanishes as $J(\equiv 1-\rho^6)$, while the one corresponding to the
second solution blows up as $1/J$. Therefore, it is reasonable to
impose $C^{(0)}_{2n}=0$ as a boundary condition. Thus, deep in the
trans-Planckian region, the field $\tilde\chi^{(0)}_1$ starts from an
extremely small value and increases with time as the momentum is
red-shifted. At the boundary between regions 0 and I, $\omega_{total}$
turns real and the solution starts oscillating. Since
$\tilde\chi^{(0)}$ is real and has only one branch, $\tilde\chi^{({\rm
I})}$ is also real and oscillates as a cosine function. The exact
solution is given in subsection 5.4. Note that the damping effect of
the $J$-dependent term has fixed the ``initial state'' in region I in
terms of the solution in region 0. For the Unruh dispersion relation,
this is no longer the adiabatic vacuum used in the literature. The
issue of the initial state dependence will be discussed in more detail 
in the next subsection. 

The behavior of $\omega_{total}$ with $v$ could change appreciably
from the one depicted in Figure 3 for other dispersion relations. For
the Unruh dispersion relations (\ref{unruh-gamma}), the length of
region I decreases with increasing $\gamma$. Another interesting case
is the dispersion relation corresponding to (\ref{soft}), depicted as
``soft'' in Figure 1. In this case, the $J$ dependent term in
$\omega_{total}$ always remains very small and as a result in region
I, well above the Planck scale, $\omega^2_{total}\approx \rho^2\approx
1$, and there will be no region 0. Thus, the problem is very similar
to the case of Unruh dispersion relation analyzed in
\cite{Martin:2000xs}. We recall that in the approach of
\cite{Kempf:2000ac}, a modified commutation relation based on
(\ref{soft}) lead to a different and much more complicated equation of
motion.

\subsection{The initial state}

At physical momenta below the Planck scale (region II), the effect of
$J$ disappears and the dispersion relation is linear, leading to the
standard free wave solutions. The coefficients $C^{\rm (II)}_n$ are
determined by matching with the solution in the trans-Planckian
regime. In region III, below the Hubble scale the modes freeze and the
solution becomes $\tilde\chi^{\rm (III)}=C^{\rm (III)}_n\,a$. The
power spectrum of fluctuations is finally given in terms of $|C^{\rm
(III)}_n|^2$. If this goes as $n^{-3}$, one ends up with the flat
spectrum. When the Hubble parameter is time dependent, the flat
spectrum receives corrections even in the absence of trans-Planckian
modifications.  Therefore, here we concentrate on the case of constant
$H$ to isolate the purely trans-Planckian effects. The possibility of
deviations from the flat spectrum due to the modified dispersion
relation, and the choice of the initial state (boundary conditions),
has been extensively discussed in the literature
\cite{Martin:2000xs},\cite{Brandenberger:2000wr},\cite{Martin:2000bv}
- \cite{Shankaranarayanan:2002ax}. The initial state is defined in the
adiabatic, {\it i.e.}, lowest order WKB, approximation. A modification
of the flat spectrum by the choice of the initial state is then
associated with the 
breakdown of this lowest order WKB approximation. Here, we formulate
the problem in terms of a (formally) exact WKB solution and argue that
the initial state dependence is determined by whether the frequency at
the initial time is real or imaginary. The result can then be
understood in terms of the breakdown of the lowest order adiabatic
approximation. However, the two criteria are not exactly equivalent in
the sense that the violation of adiabaticity does not always lead to
initial state dependence of the power spectrum. This is not in
contradiction with \cite{Brandenberger:2001ty} and
\cite{Lemoine:2001ar} since the initial state dependence there can be
attributed to a time dependent Hubble parameter. In fact, their
results show that for a constant Hubble parameter, the initial state
dependence disappears while adiabaticity is still violated, which is
further evidence for the assertion above.

During de Sitter expansion, $v=n/a=-nH\eta$. This can be used to write
the equation of motion (\ref{eomn-2}) in terms of $v$,
\beq
\left(\,\p_v^2\,+\omega_v^2\,\right)\tilde\chi_v=0\,,
\qquad
\omega_v^2(v)=\frac{\omega_{total}^2(\eta,v)}{n^2H^2}\,.
\label{eom-v}
\eeq   
From equations (\ref{d2Ja}) and (\ref{dadeS}) one can see that
$\omega_v^2$ depends only on $v$ and not on $n$ and $\eta$
separately\footnote{This will not be the case for a time dependent
Hubble parameter. For example, for $a=\eta^\lambda \ell_0$, the
equation in terms of $v$ becomes $\left(\lambda^2 v^2 \p^2_v +
\lambda(\lambda+1)v\p_v + \eta^{2\lambda+2}\ell_0^2 \rho^2(v) -
F(v)\right)\chi =0$ which explicitly depends on $\eta$, besides $v$.}.
The discussion below applies to general $\omega_v$ and is not
restricted to the situation depicted in Figure 3. The exact solutions
of the above equation can be formally written in a WKB form 
\cite{Birrell:ix},   
\beq
\tilde\chi_{v\pm}=\frac{1}{\sqrt{2W_v}}\exp\left({\pm i\int_{v_i}^v
W_{\bar v}d{\bar v}}\right)\,,
\label{WKB-v}
\eeq
where $W_v$ is given by
\beq
W_v^2=\omega_v^2 + Q_v(W_v)\,,\qquad
Q_v(W_v)=\frac{1}{2}
\left(\frac{3(\p_v W_v)^2}{2W_v^2}-\frac{\p_v^2 W_v}{W_v}\right)\,.
\eeq
This solution is formally exact, but in practice $W_v$ is determined
iteratively in terms of $\omega_v(v)$ to some order of approximation.
Near the classical turning points, where $\omega_v$ changes sign, the
integral is to be analytically continued to the complex $v$-plane to
allow for well defined solutions. This procedure, which mixes the two
branches labeled by $\pm$, is equivalent to using matching conditions
in the more familiar form of WKB \cite{Froman} . Clearly, the
solutions $\tilde\chi_{v\pm}$ are functions of $v$ alone. Let us now
turn to the solutions $\tilde\chi_\pm(\eta,v)$ of equation
(\ref{eomn-2}). These are also given by expressions analogous to the
ones above,   
\beq
\tilde\chi_\pm=\frac{1}{\sqrt{2W_{total}}}\exp\left(
{\pm i\int_{\eta_i}^\eta W_{total}d{\bar\eta}}\right)  
\label{WKB-eta}
\eeq
where, $W_{total}=nHW_v$ and $Q_{total}=nHQ_v$. It is then evident
that the exponential is the same in terms of $v$ and $\eta$.
Therefore, the dependence of $\tilde\chi$ on $n$, $v$ and
$\eta_i=-v_i/nH$ is of the form 
\beq
\tilde\chi_\pm(n,v,v_i)=\frac{1}{\sqrt{nH}}\,\tilde\chi_{v\pm}(v,v_i).
\label{chi-n}
\eeq
Thus, for a given $v$, the extra dependence on $n$ (beyond the
$1/\sqrt{n}$) could only come from the initial state at $\eta_i$.
In particular, matching the WKB solution in region I to the solution
in region II at a fixed $v=v_f$ does not introduce extra dependence on
$n$. 

The initial state dependence of the power spectrum depends on whether
$W_{total}(\eta_i)$ is real or imaginary\footnote{One often chooses
the initial state in a region where $Q_{total}$ is small so that
$\omega_{total}$ is a good approximation to $W_{total}$ (adiabatic
approximation). Then $W_{total}$ is real when $\omega_{total}$, or
$\omega_v$, is real.}. When this quantity is real, one can pick up the
positive frequency branch $\tilde\chi_{-}$ as a generalization of the
Minkowski vacuum. This is the standard way of choosing the initial
state in such cases. Then the initial state dependence will only
appear as a phase in $\tilde\chi_{-}$ through the lower limit of the
integral in (\ref{WKB-eta}). This phase will carry over to $C^{{\rm
(III)}}_n$ and will drop out of the final amplitude. The situation
will not change if adiabaticity is violated during the evolution from
$\eta_i$ to $\eta_f$, as long as $W_{total}(\eta_i)$ and
$W_{total}(\eta_f)$ are real. This corresponds to $\omega_{total}$
having an even number of classical turning points in this range.

However, if the initial state is chosen such that the solution also
contains a small negative frequency component, then the initial state
dependence does not disappear and the final amplitude will develop an
oscillatory dependence on $v_i$ or, equivalently, on $n$ for fixed
$\eta_i$. For an explicit realization of this, see
\cite{Danielsson:2002kx}. 

When $W_{total}(\eta_i)$ is imaginary, the magnitudes, and not phases,
of $\tilde\chi_\pm$ will depend on the initial state through the lower
limit of the integral. This dependence will always carry over to
region III and will show up in the power spectrum as long as $\eta_i$
is a finite time in the past. The Corley-Jacobson dispersion relation
for $b_{m}<0$ analyzed in \cite{Martin:2000xs} falls in this class.
Note that since $W_{total}(\eta_f)$ is taken to be real,
$\omega_{total}$ will have an odd number of classical turning points
between $\eta_i$ and $\eta_f$ and hence, adiabaticity is violated at
least once during the evolution.

In any case, one way of avoiding explicit dependence on $\eta_i$ (and
the associated dependence on $n$) is to specify the initial state at
$\eta_i=-\infty$. For oscillatory solutions, ($\omega^2(\eta_i)>0$)
this implies extrapolating the validity of the equation of motion to
the very beginning of inflation, which is not a very reasonable
assumption. This will simply shift the trans-Planckian problem to
higher momenta. On the other hand, for strongly damped solutions
($\omega^2(\eta_i)<<0$) it is a natural choice. Since the solution is
damped away rapidly with decreasing $\eta$, it is enough to require
the validity of the equation of motion up to a certain time $\eta_i$,
far enough in the past, but well after the beginning of inflation. Then
as far as the solution is concerned, the initial state is effectively
fixed at $\eta_i=-\infty$. The equation of motion discussed in the
last subsection admits solutions belonging to this class. The strategy
can also be applied to the Corley-Jacobson dispersion relation with
$b_{m}<0$. Another example is the model in \cite{Mersini:2001su},
although the damping there is not very strong. Both are based on
non-monotonic dispersion relations.

\subsection{Solution in regions I and 0}

Let us now get back to our equation (\ref{eomn-2}). For the Unruh
dispersion relation (\ref{unruh-gamma}) with $\gamma=3$, one gets
$J=1-\rho^6$, $\p_vJ/J= -6v^2\rho^3$ and $\p^2_vJ/J = 18v^4(3\rho^6-1) 
-12v\rho^3$. This leads to the equation, 
\beq
\Big(\p^2_\eta + a^2\rho^2 -\left[9 v^6 (2\rho^6-1) + 2\right] 
a^2H^2 \Big)\tilde\chi = 0\,.
\label{eom-g3} 
\eeq
The last two terms are plotted in Figure 3. In regions I and 0,
above the Planck scale, we can make the approximation $\rho\approx 1$
and also drop the $2$ in the square brackets. Then, the equation in
terms of $v$, reduces to  
\beq 
\left(\p_v^2+\frac{H^{-2}}{v^2}-9v^4\right)\tilde\chi_v =0\,.
\label{bessel1}
\eeq
This has exact solutions, $\sqrt{v} I_\nu(v^3)$ and $\sqrt{v}
K_\nu(v^3)$, in terms of the modified Bessel's functions. The order 
$\nu=\sqrt{1-4H^{-2}}/6$ is a large imaginary number. Using
(\ref{chi-n}), one gets   
\beq
\tilde\chi^{{\rm (0,I)}}= C_1 \sqrt{\frac{v}{n}}\, K_{\nu}(v^3) +
C_2 \sqrt{\frac{v}{n}}\, I_{\nu}(v^3)\,.
\label{solex}
\eeq
$C_1$ and $C_2$ are constants that are to be fixed by the boundary
conditions in region 0. To this end, let us consider the asymptotic
behaviour of the modified Bessel functions as $v$ goes to infinity,
\beq
K_{\nu}(v^3)\approx \sqrt{\frac{\pi}{2 v^3}}\,e^{-v^3}\,,\qquad
I_{\nu}(v^3)\approx \sqrt{\frac{1}{2\pi v^3}}\,e^{v^3}\,,
\qquad ({\rm as}\,\,\,\, v\rightarrow \infty)\,.
\eeq
Comparing (\ref{solex}) with (\ref{sol01}) and (\ref{sol02}) in this
limit, one can fix the normalizations $C^{(0)}_{1,2n} = n^{-3/2}
C^{(0)}_{1,2}$, as promised. As argued there, the growing solution
gives rise to an energy density that diverges as $v$ increases.
Therefore, we impose $C_2=0$ as a boundary condition. The surviving
solution decays rapidly with increasing $v$. As argued in the previous
subsection, this allows us to regard the initial state in the far past
as effectively corresponding to $\eta_i=-\infty$. The absence of an
explicit $\eta_i$ dependence insures the flatness of the power
spectrum of the associated fluctuations.

Thus in this scenario, in the beginning of inflation, the oscillations
are very heavily damped by the Jacobian factor $J$ and the field
$\tilde\chi$ is essentially zero. The expansion of the universe
reduces $J$ and the field starts growing in region 0. At the boundary
of regions 0 and I, $\omega_{total}$ turns positive and the solution
starts oscillating in region I. It remains real and behaves as a
cosine function with its amplitude decreasing with time. The
sinusoidal nature of the solution can be directly inferred from the
application of the WKB connection formulae at the turning point. The
solution becomes less accurate as we approach region II and can be
matched to $\tilde\chi^{(\rm II)}=C^{{\rm (II)}}_{1n}
e^{in\eta}+C^{{\rm (II)}}_{2n} e^{-in\eta}$ and its first derivative
at $v=v_p=1$. The reality of the solution in region I implies that
$C^{{\rm (II)}*}_{1n}=C^{{\rm (II)}}_{2n}$, so that $\tilde\chi^{{\rm
(II)}}=|C^{{\rm (II)}}_{n}|\cos(n\eta +
\varphi)$. The amplitude and the phase $\varphi$ are determined by the
matching and their explicit expressions can be easily worked out in
terms of $K_\nu(v^3)$ and its derivatives. Finally, $C^{\rm (III)}_n$
is determined by matching across the boundary of regions II and III.
Clearly, its dependence on $n$ is given by $n^{-3/2}$, leading to a
flat spectrum.

Note that in region 0 the field started in a WKB ground state (albeit
with an imaginary frequency). But in region I the solution behaves as
a cosine function, and is no longer in a WKB ground state. Ignoring
the issue of negative frequencies, this may be interpreted as particle
creation caused by the breakdown of adiabaticity near the classical
turning point, at the boundary between the two regions. One could
argue, however, that the back reaction problem is avoided because the
energy density (\ref{edensity}) for these modes is heavily suppressed
by J. One may also note that the solution in region III belongs to the
same branch as that in region 0 and therefore is in ground state with
respect to it. This may be taken to suggest that there is no net
particle creation in region III, with respect to region 0.

The appearance of imaginary frequencies in region 0 is one of the
features of models with an Unruh dispersion relation and a
trans-Planckian damping factor. A consequence of this is the
appearance of real fluctuations in regions I and II. In region II, for
example, this leads to $|C^{{\rm (II)}}_{1}|^2-|C^{{\rm
(II)}}_{2}|^2=0$ as opposed to $1$, which is the usual Wronskian
condition. As pointed out in the literature, the quantum field theory
of modes with imaginary frequencies is not well understood. For a
discussion, however, see \cite{Fulling:nb} and \cite{Kang:1996sx}. 

In this section we have focused on the behaviour of equation
(\ref{eomn-2}) with the dispersion relation (\ref{unruh-gamma}) for
$\gamma=3$. For other dispersion relations the behaviour could be
drastically different. As an example, the case of the dispersion
relation based on  (\ref{soft}) was discussed at the end of subsection
5.2.  

\section{Conclusions}

It is believed that fundamental physics at the Planck scale leads to a
minimum length uncertainty principle. Such a principle can in turn be
associated with a modified space-momentum commutation
relation\footnote{An alternative approach is based on space-time
uncertainty relations
\cite{Chu:2000ww},\cite{Alexander:2001dr},\cite{Lizzi:2002ib},\cite{Brandenberger:2002nq}
which is related to non-perturbative features of string theory.}. We
investigate a class of such commutators and show that the minimum
length principle can be consistently incorporated into a field theory,
both in flat space-time, as well as in the expanding universe. Planck
scale physics modifies the linear dispersion relation to a non-linear
one. Since the modification of the dispersion relation is related to
that of the commutators, it is constrained by the requirement of
minimum length. Of late, modified dispersion relations have been used
in the literature to mimic the effects of trans-Planckian physics.
However one can now see that not all of them are consistent with the
minimum length principle. Our resulting field theory can also be
understood as one in which higher-derivative corrections due to Planck
scale physics have been summed up in a closed form. Another feature of
this theory is the reduction in the number of degrees of freedom
at high energies. It is interesting to investigate the quantization of
such theories, especially when they are extended to include gravity,
since they seem to be free of ultraviolet divergences. Also, the
modified commutator theory leads to a (dynamical) breakdown of Lorentz
invariance due to the emergence of a minimum length scale, although
the background we finally consider is itself not invariant. 
On Lorentz invariance at Planck scale, especially in the context of
cosmology see, for example, \cite{Jacobson:1991gr}.

In an expanding universe, the physical momenta undergo a red-shift
which makes the non-linear dispersion relation time dependent. The
red-shift also induces a trans-Planckian damping factor in the
equation of motion. This is due to the reduction in the number of
degrees of freedom at ultra high momenta. This factor tends to
suppress the field fluctuations at trans-Planckian momenta, although
the actual effectiveness of the damping depends on the detailed form
of the dispersion relation. The damping is strong in the example of
the Unruh dispersion relation that we have considered, but can be
insignificant in some other cases. The energy density of the
fluctuations is also strongly suppressed at large momenta by the
trans-Planckian damping factor, thereby avoiding the problem of back
reaction.

The modified theory, which incorporates Planck scale physics, is then
used to study the evolution of cosmological perturbations. It is shown
that for de Sitter inflation it still leads to a flat power spectrum
for the fluctuations. In principle, the non-linear form of the
dispersion relation could complicate the choice of the initial state
whose identification at some initial time $\eta_i$ introduces
an extra scale dependence, affecting the spectral tilt. We have
discussed this in some detail and argued that, for de Sitter
inflation, the main criterion is the real or imaginary nature of the
frequency at the initial time. However, when the oscillations are
strongly damped in the far past, as is the case here, then one can
{\it effectively} treat the initial time as $\eta_i=-\infty$, even
though inflation is not really required to be eternal in the past. This
avoids the extra scale dependence and keeps the spectrum flat.

\vspace{.5cm}
\noindent{\Large{\bf Acknowledgment}}
\vspace{.3cm}
\par\noindent 
We would like to thank Kari Enqvist, Esko Keski-Vakkuri, Jens
Niemeyer, Syksy R\"{a}s\"{a}nen and Riccardo Sturani for useful
conversations and Achim Kempf for an e-mail exchange.

\end{document}